\documentclass[9pt]{amsart}

\newif\ifPreprint \Preprinttrue
\newif\ifSubmission \Submissionfalse

\usepackage{babel}
\usepackage[utf8]{inputenc}
\usepackage[T1]{fontenc}
\usepackage{lmodern}
\usepackage{xspace}
\usepackage{mathtools}
\usepackage{amsmath}
\usepackage{enumitem}
\usepackage{accents}
\usepackage{todonotes}
\usepackage{csquotes}
\usepackage{microtype}
\usepackage[citestyle=alphabetic, 
            bibstyle=alphabetic, 
            firstinits = true,
            maxbibnames = 100,
            maxcitenames = 2,
            isbn = false,
            backend = biber]{biblatex}
\usepackage{hyperref}
\hypersetup{colorlinks,
            citecolor=blue,
            urlcolor=blue,
            linkcolor=blue}
\usepackage{cleveref}

\usepackage[a4paper,top=2cm,left=2cm,right=2cm,bottom=2cm]{geometry}
\usepackage{amssymb}
\usepackage{hyperref}
\usepackage{todonotes}
\usepackage{amsthm, mathtools}

\usepackage[T1]{fontenc}
\usepackage[utf8]{inputenc}
\usepackage{color}
\usepackage{graphicx}
\usepackage{algorithm}
\usepackage{algpseudocode}
\usepackage{algorithm}
\usepackage{cleveref}
\usepackage{bbm}
\usepackage{gensymb}
\usepackage{subcaption}
\usepackage{booktabs}

\makeatletter
\patchcmd{\@settitle}{\uppercasenonmath\@title}{\scshape\large}{}{}
\patchcmd{\@setauthors}{\MakeUppercase}{\scshape\normalsize}{}{}
\makeatother

\vfuzz \hfuzz
\algdef{SE}[SUBALG]{Indent}{EndIndent}{}{\algorithmicend\ }%
\algtext*{Indent}
\algtext*{EndIndent}

\algblock{Input}{EndInput}
\algnotext{EndInput}
\algblock{Output}{EndOutput}
\algnotext{EndOutput}

\theoremstyle{plain}

\theoremstyle{definition}

\theoremstyle{remark}
\newcommand{\R}{\mathbb{R}}

\usepackage{mathrsfs}
\usepackage{xspace}
\usepackage{csvsimple}

\definecolor{cream}{RGB}{222,217,201}

\DeclareMathAlphabet{\mathcall}{OMS}{cmsy}{m}{n}
\newcommand{\secref}[1]{\hyperref[#1]{Section~\ref*{#1}}}
\newcommand{\corref}[1]{\hyperref[#1]{Corollary~\ref*{#1}}}
\newcommand{\theref}[1]{\hyperref[#1]{Theorem~\ref*{#1}}}
\newcommand{\lemref}[1]{\hyperref[#1]{Lemma~\ref*{#1}}}
\newcommand{\figref}[1]{\hyperref[#1]{Figure~\ref*{#1}}}
\newcommand{\figsref}[1]{Figures\hyperref[#1]{~\ref*{#1}}}
\newcommand{\tabref}[1]{\hyperref[#1]{Table~\ref*{#1}}}

 \newcommand{\Rho}{\mathcall{P}}
 \newcommand{\Tau}{\mathcall{T}}

\usepackage{xcolor}
\usepackage{pgfplots}
\pgfplotsset{width=10cm,compat=1.9}

\usepgfplotslibrary{external}
\bibliography{bibliography}

\begin{document}
\title{Improving reconstructions in nanotomography for homogeneous materials via mathematical optimization$^\dag$}

\author[S. Kreuz et. al.]%
{Sebastian Kreuz, Benjamin Apeleo Zubiri$^{\ast}$, Silvan Englisch, Sung-Gyu Kang, Rajaprakash Ramachandramoorthy, Erdmann Spiecker, Frauke Liers, and Jan Rolfes$^{\ast}$}

\address[S. Kreuz, F. Liers, J. Rolfes]{
Department of Data Science\\
 Friedrich-Alexander-Universität Erlangen-Nürnberg\\
 Cauerstr. 11, 91058 Erlangen, Germany\\
 Email: jan.rolfes@fau.de\\}

\address[B. Apeleo Zubiri, S. Englisch, E. Spiecker]{Institute of Micro- and Nanostructure Research (IMN)\\ $\&$ Center for Nanoanalysis and Electron Microscopy (CENEM)\\
 Department of Materials Science and Engineering\\
 Friedrich-Alexander-Universität Erlangen-Nürnberg\\
 Cauerstr. 3, 91058 Erlangen, Germany\\
 Email: benjamin.apeleo.zubiri@fau.de\\}

\address[S.-G. Kang, R. Ramachandramoorthy]{Department Structure and Nano- / Micromechanics of Materials\\
 Max-Planck-Institut für Eisenforschung GmbH\\
   Max-Planck-Str. 1, 40237 Düsseldorf, Germany\\
 Email: r.ram@mpie.de\\}

\address[S.-G. Kang]{Department of Materials Engineering and Convergence Technology\\
 Gyeongsang National University\\
   Jinju-daero 501, 
  52828 Jinju, Republic of Korea\\}

\twocolumn[\begin{@twocolumnfalse}

\maketitle


\begin{abstract}
    Compressed sensing is an image reconstruction technique to achieve high-quality results from limited amount of data. In order to achieve this, it utilizes prior knowledge about the samples that shall be reconstructed. Focusing on image reconstruction in nanotomography, this work proposes enhancements by including additional problem-specific knowledge. In more detail, we propose further classes of algebraic inequalities that are added to the compressed sensing model. The first consists in a valid upper bound on the pixel brightness. It only exploits general information about the projections and is thus applicable to a broad range of reconstruction problems. The second class is applicable whenever the  sample material is of roughly homogeneous composition. The model favors a constant density and penalizes deviations from it. The resulting mathematical optimization models are algorithmically tractable and can be solved to global optimality by state-of-the-art available implementations of interior point methods. 
In order to evaluate the novel models, obtained results are compared to existing image reconstruction methods, tested on simulated and experimental data sets. The experimental data  comprise one 360° electron tomography tilt series of a macroporous zeolite particle and one absorption contrast nano X-ray computed tomography (nano-CT) data set of a copper microlattice structure. The enriched models are optimized quickly and show improved reconstruction quality, outperforming the existing models. Promisingly, our approach yields superior reconstruction results, particularly when information about the samples is available for a small number of tilt angles only. 
\end{abstract}

\end{@twocolumnfalse} \vspace{0.6cm}]

\section{Introduction}
\label{sec:introduction}
\begin{figure}
	\centering
    \vspace{2cm}
		\begin{minipage}[c]{\linewidth}
			\includegraphics[scale=0.22]{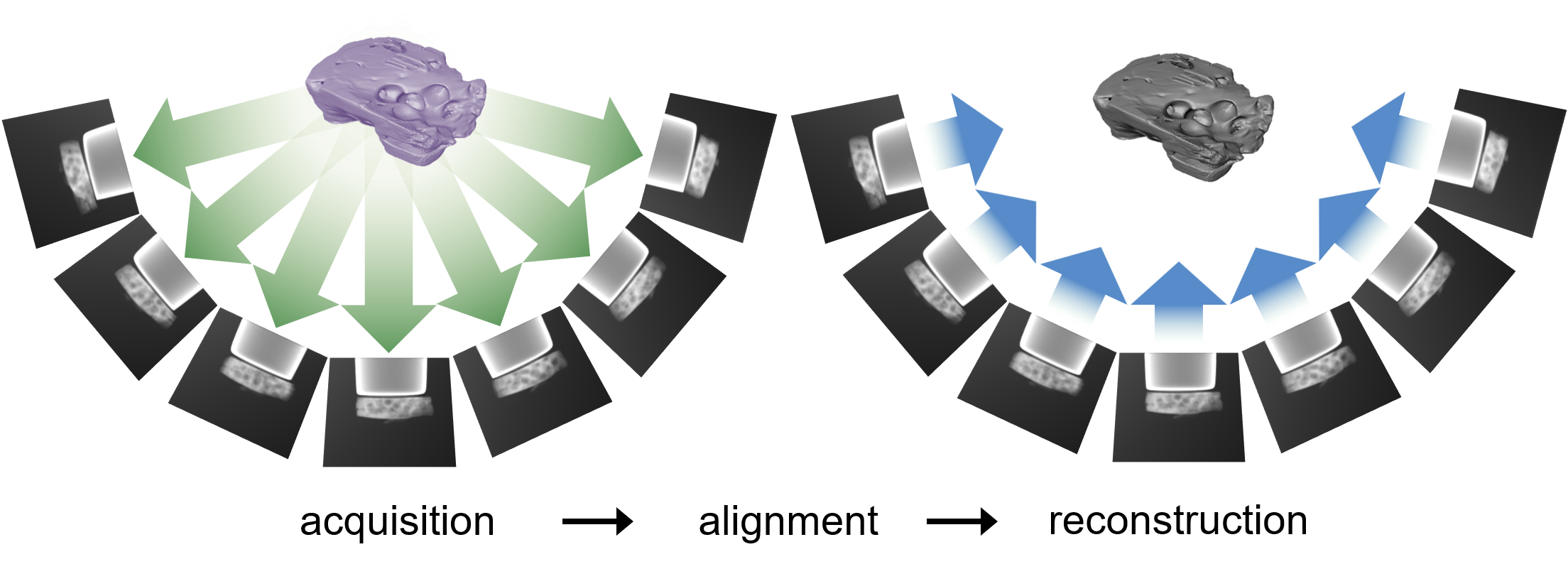}
		\end{minipage}
	\caption{Principle of projection-based nanotomography: a tilt series of projections from an object is acquired and, after an alignment with respect to the common tilt axis, the reconstruction of the object is calculated.}
	\label{tomo}
	\end{figure}

    Projection-based nanotomography techniques, including electron tomography (ET), nano X-ray computed tomography (nano-CT) and micro X-ray computed tomography (micro-CT), are designed to gain three-dimensional (3D) information from a series of two-dimensional (2D) projections of an object on a nm to mm scale.\cite{apeleo2021correlative,withers2021x,burnett2014correlative} The object, called sample or specimen, is placed on a sample holder between a typically stationary X-ray or electron beam source and a detector array in a transmission electron microscope or X-ray microscope. The projections encode information on interaction of the initial beam while penetrating through the sample. To obtain easily interpretable and reconstructable contrast, the measured intensities should exhibit a monotonic relationship to specific sample properties such as local density (i.e., mass attenuation) and thickness. In ET, this can be realized by using the so-called high-angle annular dark-field (HAADF) scanning transmission electron microscopy (STEM) imaging mode \cite{van2012correction}, whereas in nano- and micro-CT measurements, this is the case using the absorption contrast mode \cite{white2019correlative,takeya2020x}. In all cases, the sample holder or stage is tilted to perform projections from different tilt angles of the object. A visual representation of this process is depicted in \figref{tomo}.
Mathematically, the projections can be modeled as an idealized, noiseless system of linear equations
    \begin{align}
        Rf = p \label{Rf=p},
    \end{align}
    where the vector $f \in \mathbb{R}^n$ denotes the pixel brightness, $p \in \mathbb{R}^m$ the projection data and the matrix $R \in \mathbb{R}^{m \times n}$ a discretized variant Radon transform. To be more precise,    
    each row of $R$ corresponds to a single projection ray. An entry of a row displays how much of a pixel is passed by the ray in relation to its side length. Multiple projection geometries including parallel, fan, and cone beams can be modeled in this way \cite{zhuge2015tvr}. However, as \eqref{Rf=p} describes an idealized setting, introducing uncertainties, such as noise, potentially render \eqref{Rf=p} infeasible.
    Therefore, image reconstruction is usually performed via error minimization:
    \begin{align}
        \min\limits_{f} \, \lVert Rf - p\rVert_2^2 \label{minlimited},
    \end{align}
    which, due to the convexity of the norm $\|x\|_2:=\sqrt{\sum_{i=1}^n x_i^2}$, can be solved efficiently to global optimality with, e.g., gradient-based methods. In this work, we focus on the realistic and very natural setting where only few projections of the sample can be acquired. This is true, e.g.,\ if projections are expensive, time-consuming or if beam-sensitive materials are studied, which can only withstand a certain X-ray or electron dose. Under these conditions, it is crucial to take only a small number of projections, the fewer the better. Few projections lead to $m << n$. As a result, the linear equation system \eqref{Rf=p} is highly underdetermined and, consequently, the information is highly undersampled. \\
    Established iterative reconstruction techniques to solve \eqref{minlimited} 
    are, e.g., the simultaneous algebraic reconstruction technique (SART) \cite{slaney1988principles} and the simultaneous iterative reconstruction technique (SIRT) \cite{gilbert1972iterative,kak2001principles}. \\    
    Compressed sensing (CS), also named compressive sensing or compressive sampling, is a thoroughly studied signal processing technique to achieve improved reconstructions from such undersampled information by using prior knowledge about the reconstructed signal. In particular, considering the potential sparsity of the underlying signal in a known transform domain can crucially improve the reconstruction as was demonstrated in the seminal paper of Cand{\`e}s et al. \cite{candes2006stable}. CS has a wide range of applications, see Qaisar et al.\ \cite{qaisar2013compressive}, one of them is nanotomography. In that field, compressed sensing electron tomography (CS-ET) \cite{leary2013compressed} and total variation regularized discrete algebraic reconstruction technique (TVR-DART)\cite{zhuge2015tvr} are two exemplary established CS-based algorithms. The mathematical reconstruction scheme of CS is described in \secref{sec:methodology}. This work aims at improving this scheme by incorporating prior knowledge about the sample as additional constraint classes. The goal is to reduce artifacts and achieve better reconstruction quality in case of undersampled information. Two such classes of constraints are individually derived and explained in \secref{own-constraints}. On the one hand, they consist in the algebraic formulation of strict upper bounds for pixel intensities that eliminate artifacts outside of the reconstructed samples. On the other hand, we also introduce soft upper bounds for pixel intensities with which deviations above a certain value can be penalized. This latter constraint class is applicable to homogeneous samples that are composed of only one material density and exploits this problem-specific fact to increase reconstruction quality. Hence, the CS algorithm including these additional constraints will be referred to as compressed sensing for homogeneous materials (CSHM). For an optimal exploitation of these constraints, the parameter selection is crucial and is therefore illustrated in \secref{sec:implementation} along with our benchmark algorithms. To evaluate the practical impact of the new constraint classes, they are applied to one simulated and two experimental data sets, one acquired with electron tomography, the other with nano-CT (see \secref{sec:exp-details}). The results are shown in \secref{sec:compresults}, where comparisons to existing state-of-the-art algorithms are performed. It turns out that the new models lead to high-quality reconstructions within short time when compared to existing models, in particular, if only a few number of projections are available. A discussion of the results and some future extensions and applications are given in \secref{sec:concandoutl}.
\section{Compressed Sensing Framework}
\label{sec:methodology}

The following description of the CS framework is based on Leary et al.\ \cite{leary2013compressed}.
The first major assumption in CS is that there exists a known basis transformation $\psi\in \R^{n\times n}$ of the vector $f$, that is to be recovered, to a vector $c\in \R^m$ with $m << n$ nonzero coefficients $c_i$. In this case, we say that the image $f$ is \emph{compressible} in $\psi$. Consequently, it suffices to only gather information in the compressed form, i.e., instead of recording samples of $f$ directly, one records well-chosen samples $b_i$ of linear combinations $b_i=\sum_{j=1}^n \phi_{ij}f_j$. Here, $\phi\in \R^{m\times n}$ denotes known sampling coefficients, which are often inherent to the respective application, e.g., in a tomographic experiment, the measurements $b$ denote the value of line integrals with respect to projected lines through the sample, whereas the lines itself are defined by $\phi$. 

The second assumption in CS is that in order to recover $f$ efficiently, one requires the matrices $\psi$ and $\phi$ to be rather dissimilar. Similar matrices lead to redundant constraints $c=\psi f, b= \phi f$ and add little value by the measurements $b$ as most of the coefficients $b_i$ are close to zero. To this end, one measures the dissimilarity or \emph{incoherence} of $\psi$ and $\phi$. In the present paper, as well as in other nanotomography applications, incoherency with respect to the sensing matrix $R$ and the transforming matrix $\psi$ is controlled by the measuring process. Using, e.g.,\ HAADF-STEM imaging, incoherency is ensured by the independence of the different scanning position measurements \cite{hartel1996conditions}. With these two requirements, a reconstruction scheme, enforcing both sparsity in the transform domain and data consistency with the measurements, can be formulated. In particular, given the $l_p$ norm for $x\in \R^k, p \geq 1$ defined by $$ \lVert x \rVert_p := \left( \sum\limits_{i=1}^k \lvert x_i \rvert^p \right)^{\frac{1}{p}},$$
    we focus on the \emph{Basis pursuit denoising} (BPD) model introduced by Chen et al.\ \cite{chen2001atomic} for image reconstruction. For a broader overview on BPD as well as other image reconstruction models, we refer to the seminal book of Foucart and Rauhut \cite{foucart2013invitation}. In BPD, data consistency is enforced by penalizing deviations in the objective function, where $\lambda \geq 0$ weighs the sparsity
            \begin{align}
                \min\limits_{f \geq 0} \quad  \lVert Rf - p \rVert_2^2 + \lambda \lVert \psi f \rVert_1  \label{lambda-form}.
            \end{align}
    Measuring the sparsity of $\psi f$ directly via the non-convex $l_0$ norm, defined by $\lVert x \rVert_0 := \lvert \{ i: x_i \neq 0 \} \rvert$, generally leads to a more complex problem \eqref{lambda-form}. Instead, the $l_1$ norm is applied, to ensure sufficient sparsity, ideally restricting $\psi f$ to at most as many non-zero components as the number of projections $m$ \cite{foucart2013a}. 
    The BPD problem formulation \eqref{lambda-form} is widely used in the literature, e.g., in Lustig et al. \cite{lustig2007sparse} or in Block et al \cite{block2007undersampled}. \\
    The main focus of this work lies in the reconstruction of homogeneous materials. The latter are of frequent interest in nanotomography, such as in the 3D investigation of porous supports for applications in heterogeneous catalysis \cite{wirth2021unraveling}, particle chromatography \cite{johnson2018three}, fuel cells \cite{venkatesan2017probing} or battery electrodes \cite{trogadas2014x}. For homogeneous materials, reconstruction algorithms based on combinatorial optimization models have been presented in Liers and Pardella \cite{LP11}. Since homogeneous materials consist of only one approximately constant density value (i.e., composition or material phase), they provide additional information that is known beforehand. In particular, being composed of only one density value, these reconstructions are expected to exhibit either areas of said density or empty space (typically filled with vacuum or air), with only a few sharp edges in between. This promotes sparsity in its spatial finite differences. Exploiting this sparsity with compressed sensing is called total variation (TV) minimization and is often applied in the literature, e.g.,\ by Sidky and Pan \cite{sidky2008image} or Leary et al. \cite{leary2013compressed} It uses the a priori knowledge that most pixels are surrounded by pixels of the same grey level. \\
    For a quadratic picture with $n = l^2$ pixels, the sparsifying TV transform $\psi^{TV} \in \{-1, 0, 1\}^{2(n-l) \times n}$ is indirectly defined like in Block et al.\ \cite{block2007undersampled} by:
    \begin{align}
        \lVert f \rVert_{\text{TV}} := \lVert \psi^{TV} f \rVert_1 := \sum\limits_{j = 1}^n  \lVert \nabla f_j \rVert_1 = \sum\limits_{j = 1}^n \lvert \nabla^x f_j \rvert + \lvert \nabla^y f_j \rvert \label{tv-norm-def},
    \end{align}
    where for every  $j \in \{1,\ldots , n\}$, we define
    \begin{subequations}
    \label{grad-def}
        \begin{align}
            & \nabla f_j := (\nabla^x f_j, \, \nabla^y f_j) \label{grad-def-all},\\
            & \nabla^x f_j  :=
            \begin{cases}
                f_{j + 1} - f_{j} & \text{ if $j \mod l \neq 0$},\\ 
                0 & \text{ if $j \mod l = 0$},
            \end{cases}\\
            & \nabla^y f_j :=
            \begin{cases}
                f_{j + l} - f_{j} & \text{ if $j \leq n - l$},\\ 
                0 & \text{ if $j > n - l$}.
            \end{cases}
        \end{align}
    \end{subequations}
    
    For solving the CS optimization problems, several efficient algorithms are known, such as \textit{interior point methods} (IPM), \cite{kim2007interior} TVR-DART \cite{zhuge2015tvr} and CS-ET. \cite{leary2013compressed}    

\subsection{TVR-DART}
\label{sec:TVR-DART}
   We revisit the TVR-DART algorithm, developed by Zhuge et al., \cite{zhuge2015tvr} which is a compressed sensing technique specifically developed for samples consisting of only a few different density values. We use the image reconstructions computed by this algorithm as a benchmark for solutions to \eqref{cshm-prob}. 
        It aims at combining compressed sensing and discrete tomography. Homogeneous samples in nanotomography are one of many applications. It starts with estimating parameters by means of a SIRT reconstruction. Then the following optimization problem is solved
        \begin{align}
            \min\limits_{\rho_1, \ldots, \rho_G, \tau_1, \ldots, \tau_{G-1}, f} & \quad \lVert R \, S(f, \Rho, \Tau) - p \rVert_2^2  \label{tvr-prob}\\
            & + \lambda \sum\limits_{1 \leq j \leq n} M_{\delta} \left( \nabla S( \nabla f, \Rho, \Tau) \right). \notag
        \end{align}

        Here, $\rho_1, \ldots, \rho_G$ are the unknown different material density values of the sample and $\Rho := \{ \rho_1, \ldots, \rho_G \}$. The number of density values, $G \geq 1$, is assumed to be known in advance. $\tau_1, \ldots, \tau_{G-1}$ are the thresholds between the density values ($\Tau = \{\tau_1, \ldots, \tau_{G-1}\}$), setting the turning points of when a pixel density is pushed to either the lower or the higher density value. This is done by applying a differentiable soft segmentation function $S$ to each individual pixel density $f_j$. The function smoothly drives them towards the values in $\Rho$. The optimization problem \eqref{tvr-prob} is now solved for the smooth image $S(f, \Rho, \Tau)$. Zhuge et al.\ \cite{zhuge2015tvr} chose $S$ as a smooth approximation of a piecewise constant staircase function with values in $\Rho$ and jumps in $\Tau$. Instead of using the TV-norm for the sparsity term as in usual TV minimization, a differentiable norm is suggested using a Huber loss function $M_{\delta}$ with $\delta = 10^{-4}$. 
        
        Although the formulation is natural, it comes with the computational complexity that \eqref{tvr-prob} is an unconstrained continuous however non-convex optimization problem. Zhuge et al.\ use an alternating heuristic minimization procedure to find solutions in a short amount of time, starting with an initial reconstruction produced by SIRT. 
        While TVR-DART often appears effective in practical scenarios, its authors acknowledge its divergence under specific conditions, implying the absence of a guaranteed high-quality solution. This behavior is common when aiming to solve non-convex optimization problems through local heuristics. In order to overcome these drawbacks due to nonconvexities, this work introduces convex modelling extensions that can be used for roughly homogeneous materials, enabling the utilization of efficient interior point methods.
\section{Incorporating Problem Specific Information}
	\label{own-constraints}
	In this section, we introduce model extensions 
 to increase the quality of the reconstructed images. 
 They counteract different kinds of artifacts that typically appear in image reconstructions. In the following, the terms hard and soft upper bounds are used. Hard bounds mean bounds on variables that are enforced by a constraint and cannot be violated in a solution. Soft bounds, on the contrary, are bounds on variables that are not enforced but promoted by penalizing violations in the objective function weighted by an appropriately chosen parameter.
	\subsection{Hard Upper Bound on $f$ in Image Reconstruction}
	\label{sec:hard-upper-bounds}
		With \eqref{Rf=p}, i.e., assuming perfect projections without noise, it holds that
		\begin{align}
			&& p_i = \sum\limits_{j = 1}^n R_{ij} f_j \label{upper-bound-base} && \forall 1 \leq i \leq m,
		\end{align}
        with $p \geq 0$, $R \geq 0$ and $f \geq 0$. This implies that if a pixel $j$ of image $f$ gets passed by a projection ray $i$, meaning $R_{ij} > 0$, its density value $f_j$ satisfies $R_{ij}f_j \leq \sum_{j=1}^n R_{ij}f_j =p_i$ and thus the following variable bound emerges:
		\begin{align}
			\label{h-u-b}
			&& f_j \leq \min\limits_{1 \leq i \leq m: \, R_{ij} \neq 0} \, \, \frac{p_i}{R_{ij}} && \forall 1 \leq j \leq n,
		\end{align}

		We note that on the one hand, the bound $\frac{p_i}{R_{ij}}$ is loose for rays $i$ that pass through multiple sample-containing pixels $j$. This then causes the projection value $p_i = \sum_{j=1}^n R_{ij}f_j$ to be large. However, on the other hand, it is advantageous for rays that hit no sample pixels. Indeed, they have a rather small projection value $p_i$. As a consequence, all pixels passed by such a ray are restricted by a very tight bound \eqref{h-u-b}. This nudges the algorithm towards reconstructions that correctly consider little to no material at these pixels, even if that would be beneficial for the overall error minimization. Hence, the algorithm provides reduced undersampling artifacts (e.g., streaking artifacts) outside of the reconstructed sample, assuming the sample does not fill out the whole field of view of the microscope or detector and that the sample is surrounded only by air or vacuum. This effect is observed in the results later in Section \ref{sec:compresults}. We note that the impact of \eqref{h-u-b} is dependent on the average noise level of the background -- the higher the noise level, the looser the bounds. Therefore, a mean background noise subtraction preprocessing step should be applied before reconstruction.
		
		As already stated, equation \eqref{upper-bound-base} holds for many projection geometries, making the proposed bounds \eqref{h-u-b} applicable to a broad range of different tomography image reconstruction problems.
	\subsection{Soft Upper Bound on $f$ for Homogeneous Materials}
	\label{sec:soft-upper-bounds}
        Contrary to the hard upper bound \eqref{h-u-b}, the following class of constraints describes suitable bounds for sample pixels, assuming a constant material density over the whole sample. In materials science, such homogeneous samples occur frequently. For the following bounds, we assume such a homogeneous sample with the single density value $\omega \in \mathbb{R}_{\geq 0}$. If $\omega$ is not known in advance, it can be estimated. This can be done efficiently by reconstructing the same sample at a lower resolution and calculating a mean of all (bigger) positive values.

        The idea of the constraints is to quadratically penalize the density values that exceed $\omega$. Therefore, we define an auxiliary variable vector $d \in \mathbb{R}^n$ and add the constraints
		\begin{align}
			&& &d_j \geq f_j - \omega \label{soft-constr-1} && \forall 1 \leq j \leq n,\\
			&& &d_j \geq 0 \label{soft-constr-2} && \forall 1 \leq j \leq n,
		\end{align}
		to the model. Additionally, we add the quadratic term
		\begin{align}
			\mu \lVert d \rVert_2^2 = \mu \sum\limits_{j = 1}^n d_j^2 \label{soft-obj}
		\end{align}
		to the objective function of \eqref{lambda-form}. The parameter $\mu$ has to be chosen appropriately, which is further analyzed in \secref{parameters}.\\
As a result of the quadratic formulation, the higher the reconstructed intensity of a pixel is above the density $\omega$, thus being a potential artifact, the more it is penalized. This leads to a reconstructed image with fewer large deviations above $\omega$, so that the density values tend to be comparable over the whole sample. A possible downside of this model is that projection artifacts, e.g., non-linear contrast effects like remaining Bragg contrast contributions in HAADF-STEM imaging of crystalline specimen, which otherwise would be reconstructed as a brighter spot in the material, could now be wrongly distributed onto other pixels. Especially feature edges, e.g., pore edges, are prone to receive such material, since in undersampled data, the placement of edges is not always clear. The here proposed soft constraints can thus potentially lead to decreasing pore sizes in porous specimen. However, the quadratic nature of the penalty keeps this effect small. These effects can be observed later in \secref{sec:compresults}. \\
		Next, we go one step further and discuss the usage of more advanced modelling techniques from mathematical optimization. One could argue that, if homogeneous samples contain only two pixel density values, $0$ and $\omega$, density values of $f$ in the space between $0$ and $\omega$ have to be penalized as well.
        In principle, this can be accomplished by the class of constraints 
        \begin{align}
			&& & d_j \geq \lvert f_j - \omega b_j \rvert && \forall 1 \leq j \leq n \label{bin-1},\\
			&& & b_j \in \{ 0, 1\} && \forall 1 \leq j \leq n \label{bin-2}.
		\end{align}
        Consequently, all pixel values are moved towards either $0$ or $\omega$, with small differences allowed if they lead to better data consistency or smaller spatial finite differences. Edge pixels, where only a fraction of a pixel contains material and the rest is background, which are usually present in experimental data, contradict the idea of \eqref{bin-1} and \eqref{bin-2}. However, the higher the resolution, the less proportion of such pixels exist, so this drawback can be considered as neglectable for high-resolution images.
        Despite their modelling strengths, 
        including binary variables 
        significantly increases the problem's complexity. These variable turn the problem into a mixed-integer quadratic program (MIQP), which is $\mathrm{NP}$-hard in general, i.e., it is widely expected to not be efficiently solvable. In particular, this makes the problem considerably more difficult than the original convex quadratic program. \\ 
        Thus, in the remainder of this article, we focus on \eqref{soft-constr-1} and \eqref{soft-constr-2} and postpone algorithms involving binary variables 
        to future research. 
     In the following, we thus consider model
    \begin{subequations}
        \begin{align}
            \min\limits_{f} & \lVert Rf - p \rVert_2^2 + \lambda \lVert f \rVert_{TV} + \mu \lVert d \rVert_2^2 \label{soft-obj-in-Sec5}\\ 
            & d_j \geq f_j - \omega \label{soft-constr-1-in-Sec5} && \forall 1 \leq j \leq n,\\
            & d_j \geq 0 \label{soft-constr-2-in-Sec5} && \forall 1 \leq j \leq n,\\
            &  f_j \leq \min\limits_{i \in A_j}\frac{p_i}{R_{ij}} && \forall 1 \leq j \leq n,\\
            & f_j \geq 0 && \forall 1 \leq j \leq n.
        \end{align}
        \label{cshm-prob}
    \end{subequations}
    This problem will be referred to as compressed sensing for homogeneous materials (CSHM) because it exploits the homogeneity of the sample.
\section{Algorithmic Details}
	\label{sec:implementation}
    
 In this work, we focus on two-dimensional (2D) image reconstructions. 
 For a parallel beam geometry and single tilt axis tomography experiments, 3D problems can be split into independent 2D problems, making the proposed optimization problem applicable by consecutively running it for 2D data. 

    The new proposed reconstruction scheme was implemented in Python. It utilizes the ASTRA Toolbox, developed by Van Aarle et al.\ \cite{van2015astra} and the TVR-DART Toolbox, developed by Zhuge et al. \cite{zhuge2015tvr}. For solving the CS problem \eqref{cshm-prob} with hard constraints, a state-of-the-art {primal-dual interior point method (IPM)}, or barrier method \cite{gurobi2022advanced} (version 10.0.0) is used. A main advantage of IPM consists in the fact that it can very efficiently solve large-sized, linearly constrained optimization problems with convex quadratic objective functions which is exactly the form of \eqref{cshm-prob}. 
    
	\subsection{Selection of Parameters}
	\label{parameters}
        The objective function parameters $\lambda$, the weight of the total variation, and $\mu$, the penalty for exceeding the mean density $\omega$, have to be chosen before solving \eqref{cshm-prob}.
		
        Several proposals in the literature for choosing $\lambda$, e.g.,\ by Chen et al.\ \cite{chen2001atomic} or Jin and Rao \cite{jin2010algorithms}, did not lead to satisfying results. 
        For reasons of simplicity, $\lambda$ was empirically chosen in the implementation of this work. The projection data sets show that $\lambda$ can be chosen independently of pixel size or number of projection angles and yet perform well for all 2D slices of a data set. The fact that $\lambda$ does not have to be increased with the number of projection angles or image pixels, despite the increasing error minimization term, can be interpreted in the following way. First, with more image pixels and thus higher resolution, or more information about the picture through more projection angles, the quality of images returned by only error minimization of \eqref{minlimited} increases, and the undersampling artifacts decrease. A regularizing term $\lambda \lVert f \rVert_{TV}$, which is counteracting the undersampling artifacts, is needed less and less. \\
        Furthermore, the two new constraints introduce an additional smoothing effect. The soft upper bounds penalize bright spots that show bigger densities than material density $\omega$ and push the exceeding material to other pixels. Those other pixels are lying inside the sample because the hard upper bounds enforce pixels outside of the sample to be $0$ (or nearly $0$). This smoothes the reconstructed sample. In this implementation, the parameter was empirically chosen as $\lambda = 4000$ for the ET data set of the zeolite particle and the simulated data set, and $\lambda = 3$ for the nano-CT acquisition of the copper microlattice.\\
        The parameter $\mu$ in \eqref{soft-obj-in-Sec5} specifies the weight of the penalty for exceeding $\omega$ with pixel densities $f_j$. Since more projection angles or a higher pixel count generally does not lead to an increasing error term $\|d\|_2^2$, but $\omega$-exceeding artifacts remain (in contrast to undersampling artifacts), $\mu$ has to be scaled correspondingly.\\
        In X-ray and electron microscopy imaging, $\omega$-exceeding artifacts may occur due to non-linear contrast relationships between measured intensity and the specimen's mass and thickness (i.e., local mass attenuation coefficients) caused by, e.g., non-linear mass-thickness contrast  \cite{van2012correction} or remaining Bragg scattering effects \cite{venkatakrishnan2013model,venkatakrishnan2014model} in HAADF-STEM imaging. \\
        Thus, in order to keep our penalty $\mu \|d\|_2^2$ relevant with increasing problem size and corresponding increasing reconstruction error term $\| R f - p\|_2^2$, we scale $\mu$ with the problem size as follows:
        \begin{align}
            \mu = 5 a \frac{l}{256},
        \end{align}
        where $a$ specifies the number of projection angles and $l$ the number of pixels per row of the reconstructed image. The constant factor of $\frac{5}{256}$ was empirically chosen as it produced good results in preliminary tests.

     
\subsection{Reconstruction Benchmark and Hardware}
        To evaluate the quality of the developed constraints in CSHM, it is compared to three existing reconstruction techniques, namely SIRT, CS and TVR-DART. Here, CS refers to the solution of \eqref{lambda-form} with TV minimization of \eqref{tv-norm-def} and no additional constraints, but using the same IPM as CSHM. The IPM is run with a solution tolerance of $10^{-6}$. The number of iterations that the comparison algorithms SIRT and TVR-DART are supposed to perform when reconstructing images are set to $1000$ for SIRT and $250$ for TVR-DART. The initial SIRT reconstruction, which TVR-DART uses as a starting point, also runs for $1000$ iterations. Note that TVR-DART can also terminate earlier by fulfilling a convergence stopping criterion before reaching the maximum number of iterations.\\ 
        All following image reconstructions presented in \secref{sec:compresults} were performed on a laptop computer with 16 GB RAM, an AMD Ryzen 9 5900HX with Radeon Graphics, 3301 MHz, 8 CPU cores, and 16 logical processors. For a GPU-accelerated version of SIRT, a NVIDIA GeForce RTX 3050 Laptop GPU was used. 
     
\section{Experimental Details}
\label{sec:exp-details}
\begin{figure*}
        \centering
        \begin{minipage}{0.32\linewidth}
            \includegraphics[scale=0.48]{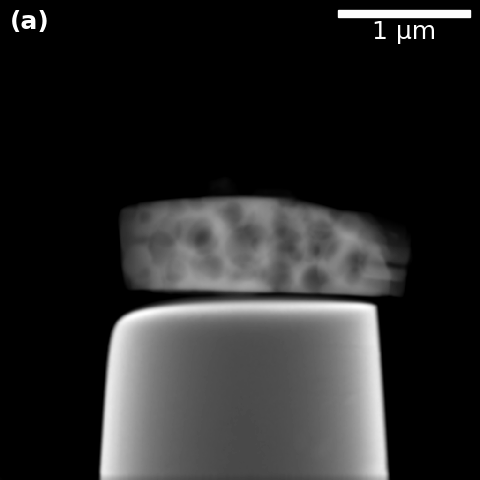}
        \end{minipage}
        \begin{minipage}{0.32\linewidth}
            \includegraphics[scale=0.48]{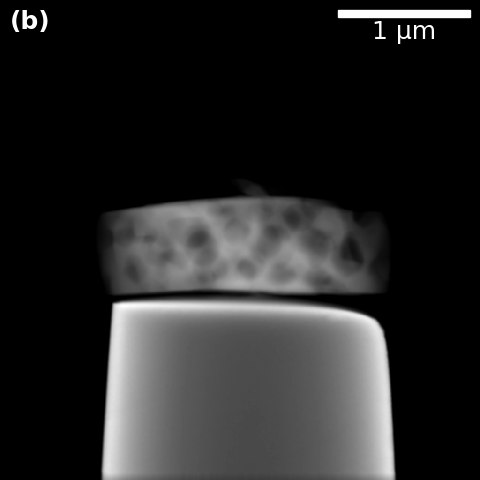}
        \end{minipage}
        \begin{minipage}{0.32\linewidth}
            \includegraphics[scale=0.48]{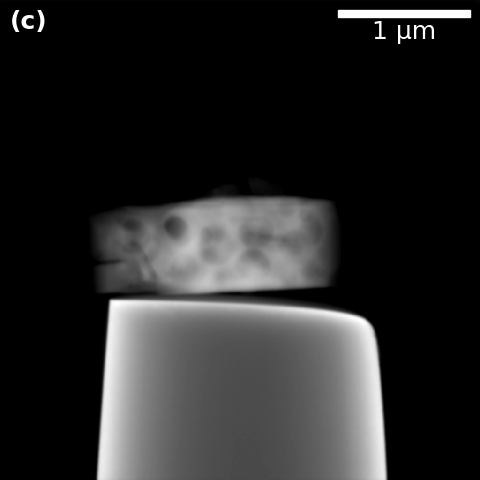}
        \end{minipage}
        \caption{Exemplary projections of experimental ET data from different projection angles of a macroporous zeolite particle on the plateau of a tomography tip: (a) 20°, (b) 90°, (c) 160°. See SI Video 1 for an animation of the full tilt series. }
        \label{proj-showing-et1}
    \end{figure*}

\begin{figure*}
        \centering
        \begin{minipage}{0.32\linewidth}
            \includegraphics[scale=0.48]{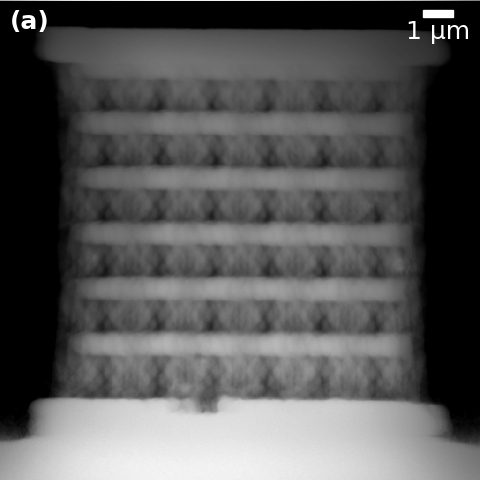}
        \end{minipage}
        \begin{minipage}{0.32\linewidth}
            \includegraphics[scale=0.48]{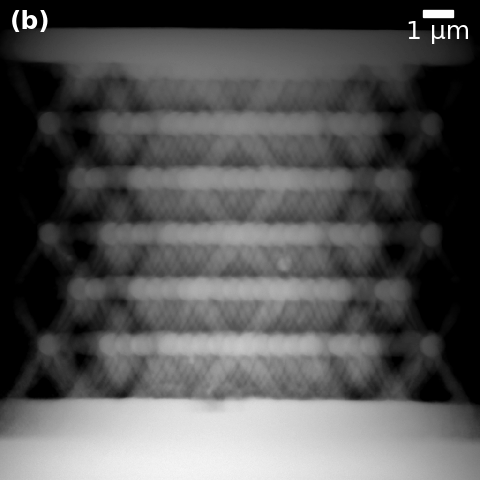}
        \end{minipage}
        \begin{minipage}{0.32\linewidth}
            \includegraphics[scale=0.48]{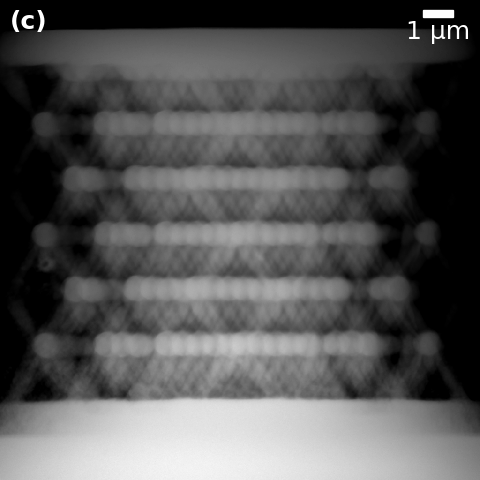}
        \end{minipage}
        \caption{Exemplary projections of experimental absorption-contrast nano-CT data set from different projection angles  of a copper microlattice structure: (a) 0°, (b) 10°, (c) 30°. See SI Video 2 for an animation of the full tilt series.}
        \label{proj-showing-et2}
    \end{figure*}

\begin{figure*}
        \centering
        \begin{minipage}{0.32\linewidth}
            \includegraphics[scale=0.48]{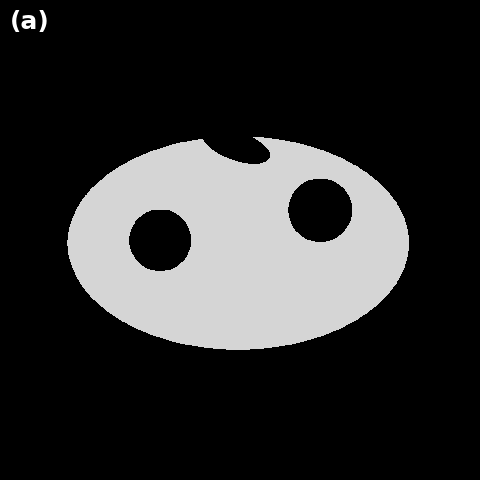}
        \end{minipage}
        \begin{minipage}{0.32\linewidth}
            \includegraphics[scale=0.48]{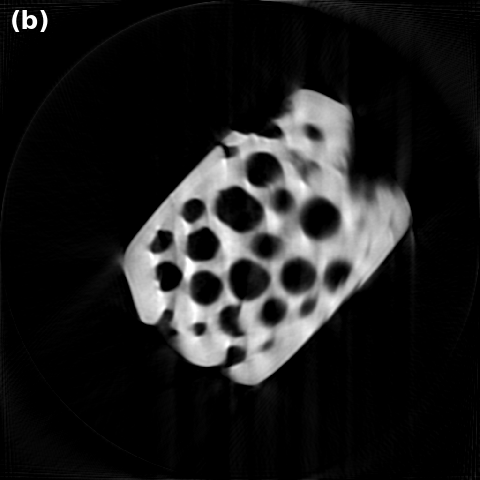}
        \end{minipage}
        \begin{minipage}{0.32\linewidth}
            \includegraphics[scale=0.48]{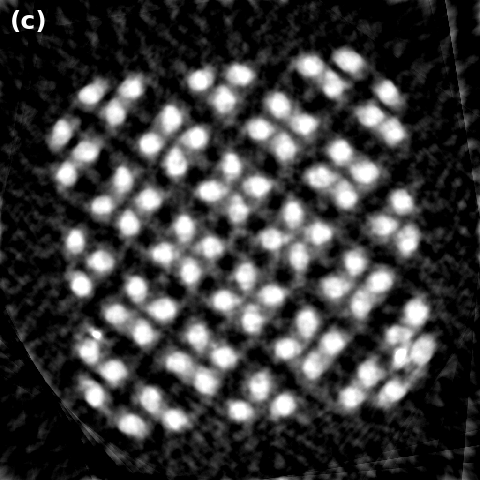}
        \end{minipage}
        \caption{"Ground truth" images for following reconstructions: (a) simulated phantom structure, (b) SIRT reconstruction with 1000 iterations using 180 projections in a 180° tilt-angle range of a macroporous zeolite particle, (c) SIRT reconstruction with 1000 iterations using 40 projections in a 180° tilt-angle range of a copper microlattice structure.}
        \label{ground-truth-image-comps}
    \end{figure*}

    Image reconstructions considered in this article are performed on three data sets: one simulated phantom object slice (\figref{ground-truth-image-comps}a) with added random Poisson noise and two experimental data sets, one acquired by ET and one by nano-CT. \\
	The ET data in \figref{proj-showing-et1} comprises a macroporous MFI-type zeolite particle synthesized by Machoke et al. \cite{machoke2015micro}. A full tilt series, as shown in SI Video 1, in a tilt-angle range of 180° with 1° tilt increment (in total 180 projections) of the particle on top of the plateau of a tomography tip (so-called 360°-ET or on-axis ET) was acquired using a FEI Titan$^{3}$ 80-300 transmission electron microscope operated at an acceleration voltage of 200 kV in HAADF-STEM imaging mode (image size 1024 pixels x 1024 pixels; pixel size 3.55 nm) and a Fischione Model 2050 On-Axis Rotation Tomography sample holder (E.A. Fischione Instruments, Inc.). HAADF-STEM imaging assures an approximately parallel beam geometry and a monotonous relationship between measured intensities, sample mass (density), and thickness, so that the Radon transform can be applied for reconstruction. For more experimental details, please refer to Przybilla et al. \cite{przybilla2018transfer}. \\
The nano-CT data set in \figref{proj-showing-et2} shows a copper microlattice. These copper microlattices were manufactured using an additive micromanufacturing technique based on localized electrodeposition in liquid (CERES system – Exaddon AG, Switzerland) \cite{kang2023fabrication, ramachandramoorthy2022anomalous}. The electrochemical ink required for the electrodeposition is supplied with a hollow atomic force microscope (AFM) cantilever with a small orifice ($\textasciitilde$300nm) and the metal ions get reduced on the conductive substrate that acts as the working electrode. Copper microlattices were built in a voxel-by-voxel manner using precise piezo positioners which move the AFM cantilever after each voxel deposition to a new location based on the deflection of the AFM cantilever sensed using a laser based optical tracking system. \\  The nano-CT experiment was performed using a ZEISS Xradia 810 Ultra laboratory-scale X-ray microscope equipped with a 5.4 keV rotating anode Cr source in absorption contrast and large-field-of-view (LFOV) mode. The LFOV mode covers a field of view (FOV) of 65 µm x 65 µm with a spatial resolution down to 150 nm and a pixel size of 63.89 nm. A tilt series with $40$ projections in a tilt-angle range of 180° (tilt increment 4.5°) and an exposure time of 300 s/frame was acquired, as shown in SI Video 2. We assume parallel beam geometry and a beam attenuation according to Beer-Lambert law, so that the Radon transform can be applied for reconstruction using the projection data $p = - \ln{I/I_{0}}$, where $I$ is the measured intensity and $I_{0}$ is the unattenuated incident beam intensity. The nano-CT projection images in \figref{proj-showing-et2} are displayed with inverted contrast. Please refer to SI Movie 2 for the animated video of the original absorption contrast nano-CT tilt series. \\
    The 360°-ET tilt series was aligned by cross correlation in FEI Inspect 3D version 3.0. The nano-CT tilt series was aligned using the adaptive motion compensation (AMC; based on Wang et al. \cite{Wang2012}) procedure implemented in the native ZEISS software (XMController).
    

    
\section{Computational Results}
\label{sec:compresults}

    As an objective quality measure for the results of the simulated data set reconstructions, the relative mean error (RME)
    \begin{align}
        \text{RME} := \frac{\sum\limits_{j=1}^n \lvert f_j - \tilde{f}_j \rvert}{\sum\limits_{j=1}^n \lvert \tilde{f}_j \rvert}
    \end{align}
    can be calculated, like in Zhuge et al.\ \cite{zhuge2015tvr}, comparing the reconstructed image $f$ and the original image without noise ($\tilde{f}$). Since there exist no ground truth images for the experimental data, the RME cannot be calculated there. Instead, one can calculate the raw data coverage (RDC) 
    \begin{align}
        \text{RDC} :=  \frac{\sum\limits_{i=1}^{\bar{m}} \lvert \bar{R} f^{*} - \bar{p} \rvert}{\sum\limits_{i=1}^{\bar{m}} \lvert \bar{p} \rvert}
    \end{align}
    that displays how well the projection of a reconstruction $f^{*}$, attained with (possibly limited) projection data $p$, aligns with all projection data $\bar{p}$ of all $\bar{m}$ available angles. $\bar{R}$ is the corresponding radon matrix. Note that the RDC is not as meaningful or reliable as the RME concerning reconstruction quality. This is due to the fact that the noise in the original projection data $\bar{p}$ is not known and therefore part of the comparison. For a visual comparison, \figref{ground-truth-image-comps} shows "ground truth" images of the 2D slices that are reconstructed later in this section. Since there are no ground truths for the experimental data, the standard image reconstruction method SIRT is applied to the whole available projection data ($1024^2$ pixels each, $180$ angles in \figref{ground-truth-image-comps}b, $40$ angles in \figref{ground-truth-image-comps}c) to attain comparison images.

	\subsection{Results}
	\label{results-individual}
		As compressed sensing is designed to solve image reconstruction problems with undersampled data, meaning only a few projection angles are available or the projection range contains a missing wedge, primarily those problem settings are studied. The projection angles in the following reconstructions are always equidistantly spread over the available angular range. That is, 0° to 179° if there is no missing wedge, and $\frac{x}{2}\degree$ to $(180-\frac{x}{2})\degree$ if there is a missing wedge of $x\degree$. 

		The CSHM technique is compared to the existing image reconstruction techniques CS, SIRT, and TVR-DART. Except for SIRT, all algorithms are specifically designed to solve problems with undersampled projection data. The evaluation is separately performed for simulated, experimental, and missing wedge projection data.
		\subsubsection*{Simulated Data}
\begin{figure*}
            \captionsetup[subfigure]{justification=centering}
				\centering
				\subcaptionbox{SIRT: $3$s, RME: $0.7220$ \label{sirt-comp-simul-1}}[.24\textwidth]{
					\includegraphics[scale=0.36]{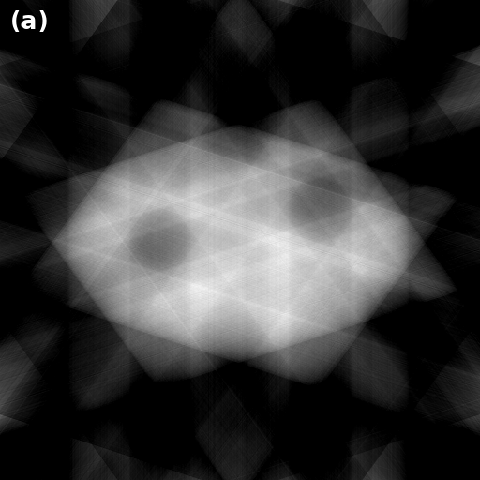}
				}
				\subcaptionbox{CS: $150$s, RME:\hspace{-1pt} $0.0755$ \label{cs-comp-simul-1}}[.24\textwidth]{
					\includegraphics[scale=0.36]{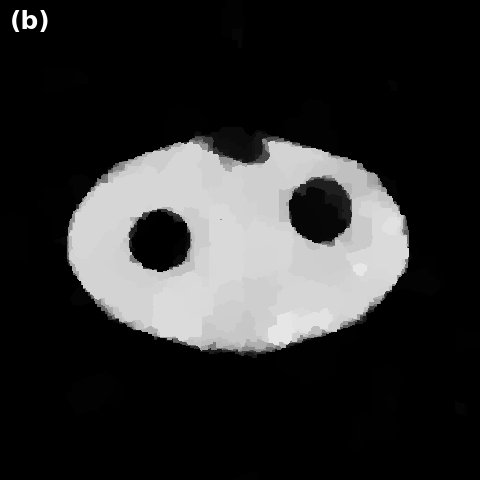}
				}
				\subcaptionbox{TVR-DART: $184$s, RME: $0.1618$ \label{tvr-dart-comp-simul-1}}[.24\textwidth]{
					\includegraphics[scale=0.36]{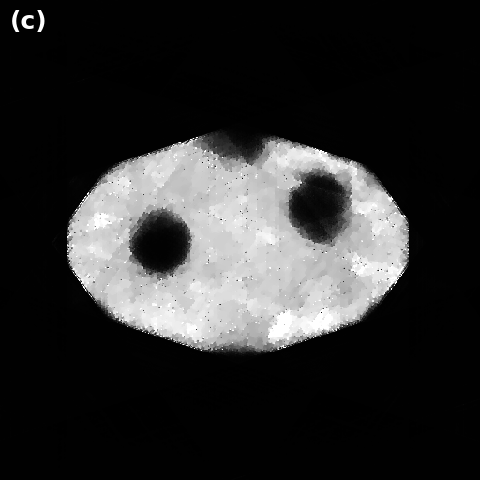}
				}
				\subcaptionbox{CSHM: $56$s, RME: $0.0413$ \label{cshm-comp-simul-1}}[.24\textwidth]{
					\includegraphics[scale=0.36]{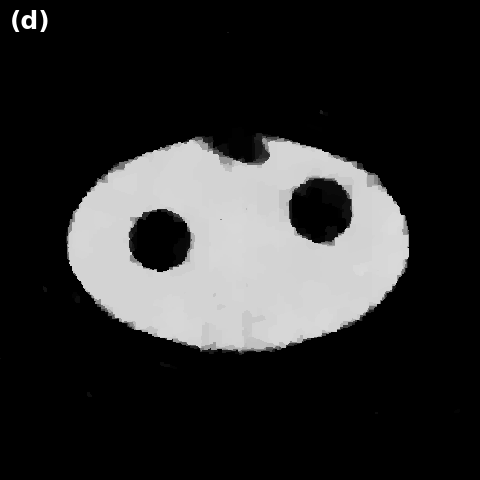}
				}
				\caption{Results for simulated data with $512^2$ pixels and $5$ projection angles.}
				\label{fig:simul-comp-all-1}
			\end{figure*}

			First, we compare reconstructions of the simulated data set with $512^2$ pixels and only $5$ projection angles (no missing wedge). The results are displayed in \figref{fig:simul-comp-all-1}, where corresponding running times and RMEs are noted below the reconstructions.
            \begin{figure*}
            \captionsetup[subfigure]{justification=centering}
				\centering
				\subcaptionbox{SIRT: $5$s, RME: $0.3209$ \label{sirt-comp-simul-2}}[.24\textwidth]{
					\includegraphics[scale=0.36]{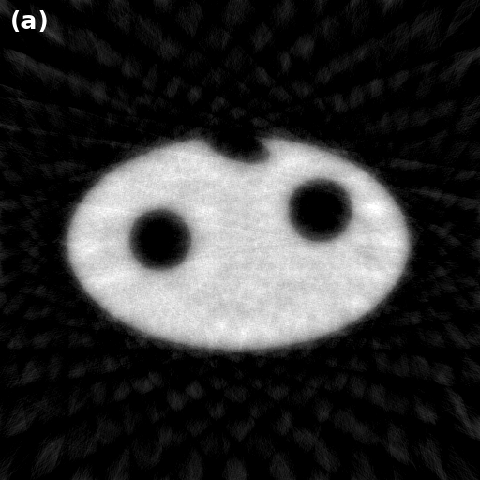}
				}
				\subcaptionbox{CS: $307$s,\hspace{-1pt} RME: $0.0525$ \label{cs-comp-simul-2}}[.24\textwidth]{
					\includegraphics[scale=0.36]{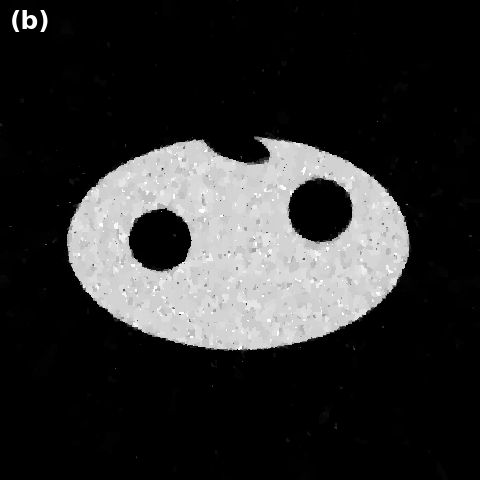}
				}
				\subcaptionbox{TVR-DART: $20$s, RME: $0.0420$ \label{tvr-dart-comp-simul-2}}[.24\textwidth]{
					\includegraphics[scale=0.36]{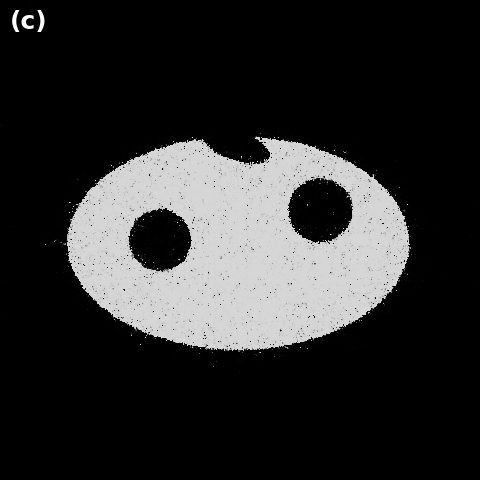}
				}
				\subcaptionbox{CSHM: $111$s, RME: $0.0301$ \label{cshm-comp-simul-2}}[.24\textwidth]{
					\includegraphics[scale=0.36]{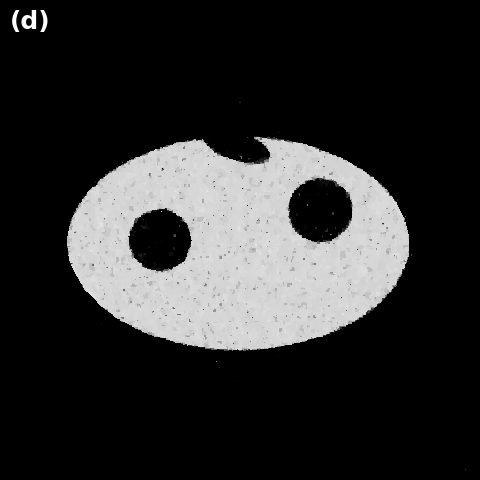}
				}
				\caption{Results for simulated data with $512^2$ pixels and $20$ projection angles.}
				\label{fig:simul-comp-all-2}
			\end{figure*}
			The small number of only $5$ angles was chosen because the shape of the object is very simple and $5$ angles are already enough for good reconstructions in CS approaches. For SIRT, however, this number of projection angles is too small. Its reconstruction in \figref{sirt-comp-simul-1} is far from accurate, which is emphasized by an RME of $0.7220$. Both CS and CSHM in \figsref{cs-comp-simul-1} and \ref{cshm-comp-simul-1} produce good results with a qualitative advantage of CSHM. This is also reflected in the relative mean errors, the RME of CSHM is about half the one of CS (0.0413 vs. 0.0755). The TVR-DART result in \figref{tvr-dart-comp-simul-1} is better than the one from SIRT, but worse than CS and CSHM reconstructions. Concerning the running times, SIRT is the fastest approach. CSHM does not only produce qualitatively better results than CS, but is also about three times faster. This is owed to the hard upper bounds eliminating a great number of variables. The TVR-DART runs the longest, which, in combination with its reconstruction quality, indicates that it did not converge and iterated until the given maximum number of iterations.

            In \figref{fig:simul-comp-all-2}, the same reconstructions are performed with $20$ projection angles. The SIRT reconstruction in \figref{sirt-comp-simul-2} is still very blurry with a large RME of $0.3209$. CSHM is again about three times faster than CS while producing an about two times better RME. Comparing these two reconstructions shows that the CS reconstruction in \figref{cs-comp-simul-2} exhibits noisy spots, both dark and bright, while the CSHM reconstruction in \figref{cshm-comp-simul-2} only shows the dark spotted noise. The brighter spots are weakened or even eliminated by the soft upper bounds by penalization. The TVR-DART reconstruction in \figref{tvr-dart-comp-simul-2} also only exhibits darker spots but it also shows a lot of spotted artifacts around the sample. Those are less for CS and nearly absent for CSHM due to the hard upper bounds for pixels outside of the sample. The RME of TVR-DART, $0.0420$, is better than the $0.0525$ for CS, but still worse than the $0.0301$ of CSHM. The much faster running time of TVR-DART compared with the reconstruction with only $5$ projection angles suggests that it converged.
			
		  \begin{table*} 
                \centering
                \small
                \caption{Results for simulated projection data with $256^2$ pixels and different amount of projection angles. The best RME value for each number of projection angles is highlighted in bold.}
                \label{result-table}
                \begin{tabular*}{\textwidth}{@{\extracolsep{\fill}}lllllllllll}
\toprule
& & \multicolumn{4}{c}{Running Time (in seconds)} & & \multicolumn{4}{c}{Relative Mean Error (RME)}\\
\cmidrule(lr){3-6}
\cmidrule(lr){8-11}
Proj. Angles      & & SIRT & CS & TVR-DART & CSHM & & SIRT & CS & TVR-DART & CSHM  \\
\midrule
5   &           & 3       & 20  & 61 & 16 &  & 0.5605 & 0.0601 & 0.1537 & \textbf{0.0274}  \\
10   &           & 3       & 30  & 8 & 19 &  & 0.3382 & 0.0411 & 0.0375 & \textbf{0.0262}  \\
15   &           & 3       & 49  & 8 & 20 &  & 0.2511 & 0.0414 & 0.0365 & \textbf{0.0269}  \\
20   &           & 3       & 57  & 4 & 22 &  &0.2124 & 0.0421 & 0.0314 & \textbf{0.0247}  \\
30   &           & 3       & 114  & 4 & 24 &  & 0.1686 & 0.0468 & 0.0282 & \textbf{0.0243}  \\
45   &           & 3       & 227  & 7 & 54 &  & 0.1390 & 0.0539 & 0.0278 & \textbf{0.0240}  \\
60   &           & 3       & 478  & 6 & 92 &  & 0.1279 & 0.0578 & 0.0236 & \textbf{0.0232}  \\
90   &           & 3       & 850  & 6 & 428 &  & 0.1200 & 0.0628 & \textbf{0.0221} & 0.0223  \\
180   &           & 3       & -  & 13 & 2897 &  & 0.1244 & - & 0.0225 & \textbf{0.0202}  \\
\bottomrule
\end{tabular*}
            \end{table*}
			To get some more insights, a comparison in RME and running time is displayed in \tabref{result-table} for a smaller pixel count of $256^2$ and different numbers of projection angles. The best RME of every row is displayed in bold. SIRT solved every problem within $3$ seconds. Both CS and CSHM exhibit strictly monotonically increasing running times, which is expected for increasing problem sizes solved with an IPM. CSHM is consistently faster than CS. The biggest problem with $180$ projection angles could not be solved anymore with CS because the computer ran out of memory. The running times of TVR-DART are very low for all reconstructions except for the one using projections from $5$ angles, indicating that the algorithm failed to converge with only $5$ angles of information but managed to do so for every other number. This claim is supported by comparing the relative RME differences between $5$ and $10$ projection angles for all algorithms, showing by far the largest for TVR-DART.
   
            Despite a monotonically decreasing RME for SIRT (except for the last value), the lowest RME of SIRT, $0.1200$, is bigger than the RME that all the other algorithms achieved with any number of projection angles, except TVR-DART with $5$ angles. The RME of CS decreases from $5$ to $10$ angles and then slightly increases in every step. This might be caused by the chosen $\lambda$ (see \secref{parameters}) not scaling well for an increasing number of projections without the additional smoothing effect by the new constraints. The RME of TVR-DART decreases in every step except for the last one where it increases slightly. Despite two small exceptions, the RME value for CSHM overall decreases with an increasing number of projection angles and reaches a minimum of $0.0202$ at $180$ angles. This is the lowest RME any of the tested algorithms could achieve for any number of projection angles.

			In eight out of nine cases, CSHM reached the lowest RME out of the four algorithms, only one times slightly outperformed by TVR-DART for $90$ projection angles. It produced a better reconstruction with only $5$ angles than CS or SIRT did for any given number of projections while maintaining a running time $\leq 60$ seconds for problems with $45$ or fewer angles. Note that the RME of CSHM acquired with $5$ projection angles is already very small and does not improve by much anymore with more projection angles. This proves the strength and high efficiency of CSHM for sparse data sets. Indeed, it yields very good and high-quality results with relatively low running times in particular in the situation when reconstructing from only a few number of projections. Having only a low number of projections at hand is a typical situation in many settings. This advantage is further fostered in the reconstructions of the following experimental data sets.
   
		\subsubsection*{Experimental Data}
			Since the experimental data sets do not have a ground truth image, the following reconstructions are only compared visually and by the RDC, a quality measure that is not robust against artifacts appearing in the projection data before reconstructing.
            \begin{figure*}
            \captionsetup[subfigure]{justification=centering}
				\centering
				\subcaptionbox{SIRT: 4s, RDC: $18.0740$ \label{sirt-comp-exp-1}}[.24\textwidth]{
					\includegraphics[scale=0.36]{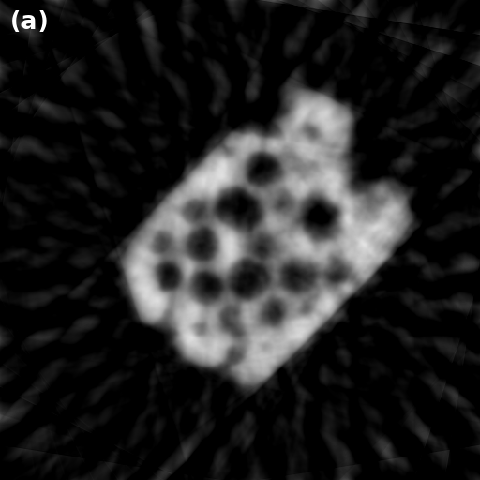}
				}
				\subcaptionbox{CS: 262s, RDC: $8.0577$ \label{cs-comp-exp-1}}[.24\textwidth]{
                    \includegraphics[scale=0.36]{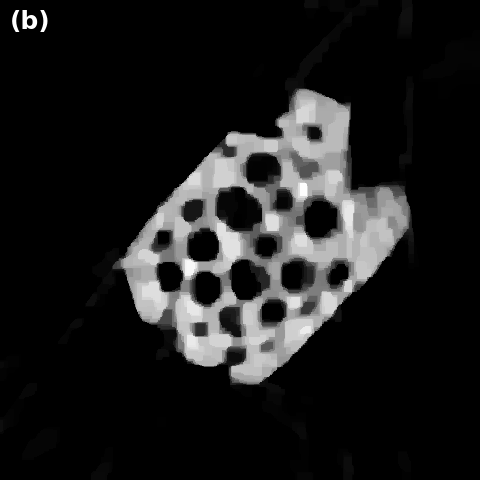}
				}
				\subcaptionbox{TVR-DART: 65s, RDC: $11.3882$ \label{tvr-dart-comp-exp-1}}[.24\textwidth]{
                    \includegraphics[scale=0.36]{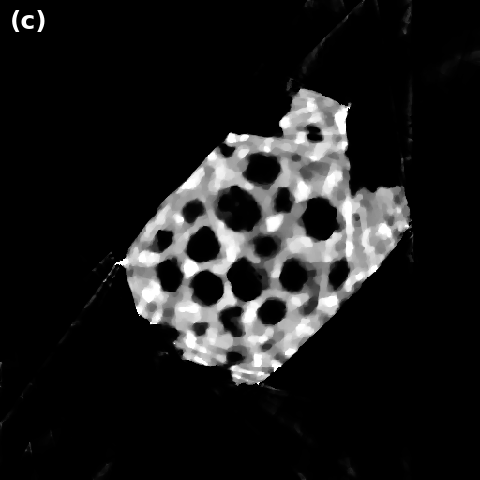}
				}
				\subcaptionbox{CSHM: 94s, RDC: $8.4071$ \label{cshm-comp-exp-1}}[.24\textwidth]{
                    \includegraphics[scale=0.36]{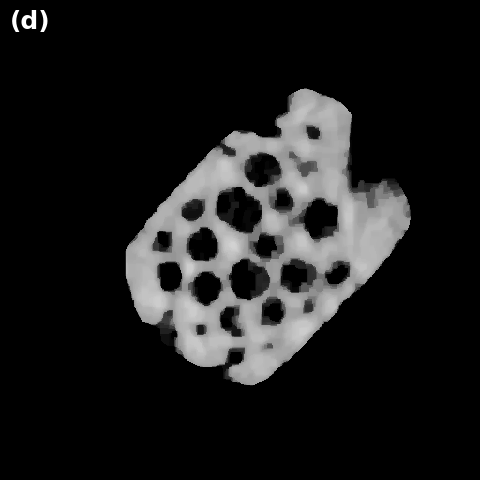}
				}
                \caption{Results for experimental data set of a zeolite with $512^2$ pixels and $20$ projection angles.}
				\label{fig:exp-comp-exp-1-label}
			\end{figure*}
            \begin{figure*}
            \captionsetup[subfigure]{justification=centering}
				\centering
				\subcaptionbox{SIRT: $4$s, RDC: $10.8665$ \label{sirt-comp-exp-1-2}}[.24\textwidth]{
					\includegraphics[scale=0.36]{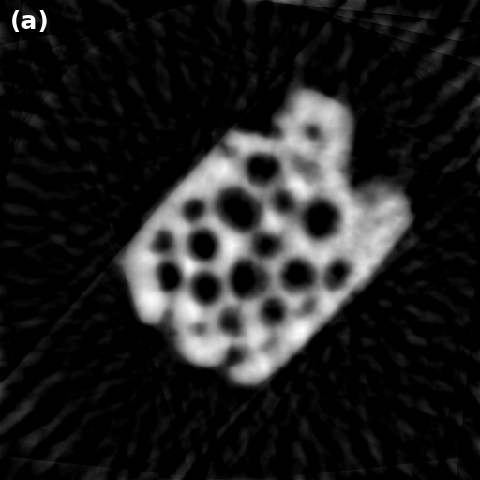}
				}
				\subcaptionbox{CS: $588$s, RDC: $7.0235$ \label{cs-comp-exp-1-2}}[.24\textwidth]{
                    \includegraphics[scale=0.36]{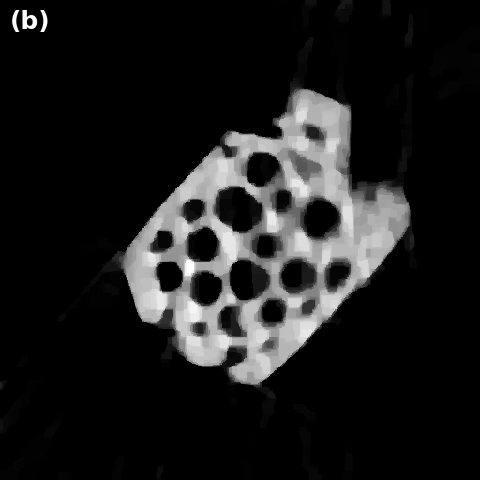}
				}
				\subcaptionbox{TVR-DART: $75$s, RDC: $9.2214$ \label{tvr-dart-comp-exp-1-2}}[.24\textwidth]{
                    \includegraphics[scale=0.36]{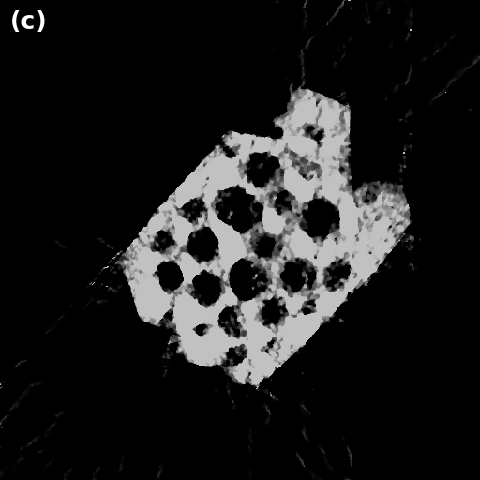}
				}
				\subcaptionbox{CSHM: $129$s, RDC: $7.8802$ \label{cshm-comp-exp-1-2}}[.24\textwidth]{
                    \includegraphics[scale=0.36]{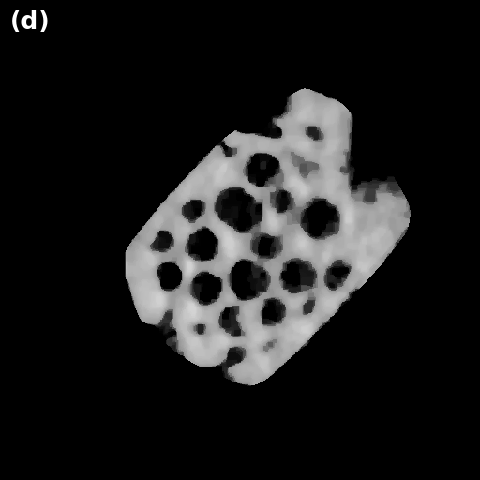}
				}
                \caption{Results for experimental data set of a zeolite with $512^2$ pixels and $30$ projection angles.}
				\label{fig:exp-comp-exp-1-1-label}
			\end{figure*}
            \begin{figure*}
                \centering
                \subcaptionbox{SIRT: 2s, RDC: $30.7225$ \label{sirt-comp-exp-2}}[.24\textwidth]{
					\includegraphics[scale=0.36]{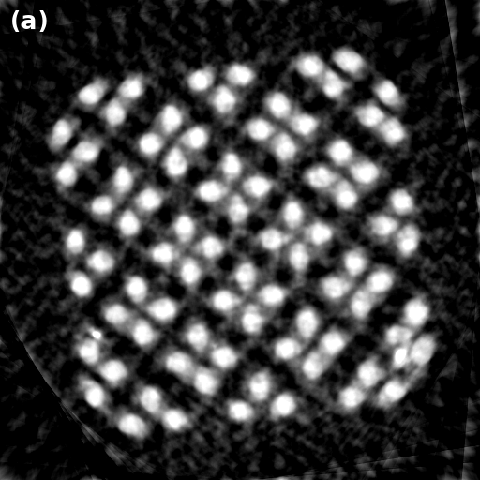}
				}
				\subcaptionbox{CS: 1215s, RDC: $31.2364$ \label{cs-comp-exp-2}}[.24\textwidth]{
                    \includegraphics[scale=0.36]{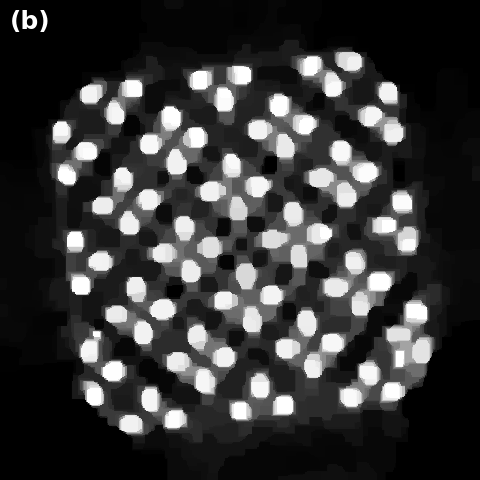}
				}
				\subcaptionbox{TVR-DART: 52s, RDC: $31.5497$ \label{tvr-dart-comp-exp-2}}[.24\textwidth]{
                    \includegraphics[scale=0.36]{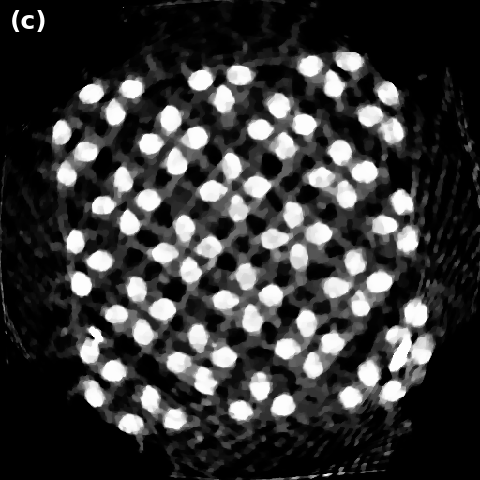}
				}
				\subcaptionbox{CSHM: 640s, RDC: $30.5611$ \label{cshm-comp-exp-2}}[.24\textwidth]{
                    \includegraphics[scale=0.36]{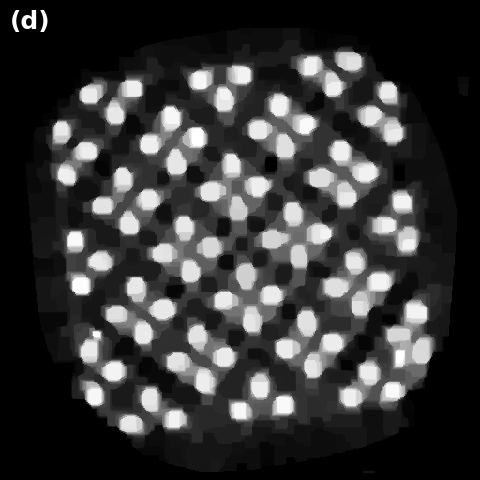}
				}
				\caption{Results for experimental data set of a Cu-microlattice with $512^2$ pixels and $40$ projection angles.}
				\label{fig:exp-comp-exp-2-label}
			\end{figure*}

            In \figref{fig:exp-comp-exp-1-label}, reconstructions of the experimental electron tomography projection data of the zeolite particle sample with SIRT, CS, TVR-DART and CSHM are displayed. The images contain $512^2$ pixels and  $20$ projection angles (no missing wedge) were used. Below the reconstructions, the corresponding running times and RDCs are noted. The running times are comparable to the ones from the simulated data set with $5$ projection angles in \figref{fig:simul-comp-all-1}. SIRT is by far the fastest algorithm, CSHM is considerably faster than CS, and TVR-DART takes relatively long, if not as long as for the simulated data set. This indicates that TVR-DART converged this time, but not very quickly. In terms of quality, the SIRT reconstruction in \figref{sirt-comp-exp-1} is very blurry and exhibits a lot of artifacts. The TVR-DART reconstruction in \figref{tvr-dart-comp-exp-1} is less blurry but shows frequently alternating brightness inside the material, which should not be the case for a homogeneous sample. The reconstruction from CS in \figref{cs-comp-exp-1} has considerably less bright spots than TVR-DART but a few of those artifacts can be observed here, as well. CSHM produces the best-looking reconstruction for the zeolite in \figref{cshm-comp-exp-1}. It displays almost no artifacts outside of the sample, due to the hard upper bounds, and has a nearly constant material density over the sample, due to the soft upper bounds. The shape of the pores in the CSHM reconstructions might be a little more compressed than in reality, as, e.g., the indentation on the upper left side of the particle in \figref{cshm-comp-exp-1} appears a little pressed together compared to the $180$ angles SIRT reconstruction in \figref{ground-truth-image-comps}b. The corresponding raw data coverage values support the visual evaluation that the SIRT and TVR-DART results are worse than CS and CSHM. However, the RDC of the CS result with $8.0577$ is slightly better than the RDC of $8.4071$ for CSHM, despite CSHM being visually the better reconstruction. This may be owed to noise during the projection data acquisition that is reconstructed with CS but penalized and lessened by CSHM. The RDC is therefore only a rough quality measure here.

            In \figref{fig:exp-comp-exp-1-1-label}, the same reconstructions are performed with $30$ projection angles. The SIRT reconstruction in \figref{sirt-comp-exp-1-2} improved a lot, exhibiting less blur and less artifacts outside of the sample. The CS and CSHM reconstructions improved only by small amounts, the pore edges are a little sharper than before. The TVR-DART reconstruction in \figref{tvr-dart-comp-exp-1-2} is very different than for $20$ angles in \figref{tvr-dart-comp-exp-1}. Now there seems to exist only one density value over the whole sample, but a lot of small black spots, pepper noise, can be found across the whole sample. The RDC values for all algorithms improved. Again, the raw data coverage value of the CS reconstruction is better than the one of the CSHM reconstruction, whereas visually, the one from CSHM is a more realistic image of a homogeneous sample. \\
            \figref{fig:exp-comp-exp-2-label} compares the different reconstructions of an exemplary slice from the copper microlattice nano-CT tilt series in absorption contrast from \figref{proj-showing-et2} using 40 projection angles. From a qualitative perspective, CSHM and CS provide the best reconstructions with no streaking artifacts in the inside of the lattice structure and especially also not in the outside regions due to the hard upper bound. The bright main copper strut features appear more homogeneous in intensity for CSHM due to the soft upper bound, with slightly higher intensity variations in the CS slice. Although TV-DART provides the better result, both SIRT and TVR-DART reconstructions show artifacts around the outside of the specimen, and, in the sample interior, remaining streaking artifacts and stronger contrast variations are observed. Due to the increased amount of projection angles, also the running time of CS and CSHM are increased, whereas SIRT is the fastest, and also TVR-DART seems to have converged rather quickly. With respect to the RDC, all four reconstructions exhibit similar values, with CSHM just showing the lowest RDC.
		\subsubsection*{Missing Wedge}
\begin{figure*}
				\centering
				\subcaptionbox{SIRT, 4s, RME: $0.5775$ \label{cs-simul-sirt-comp-missing-wedge}}[.24\textwidth]{
					\includegraphics[scale=0.36]{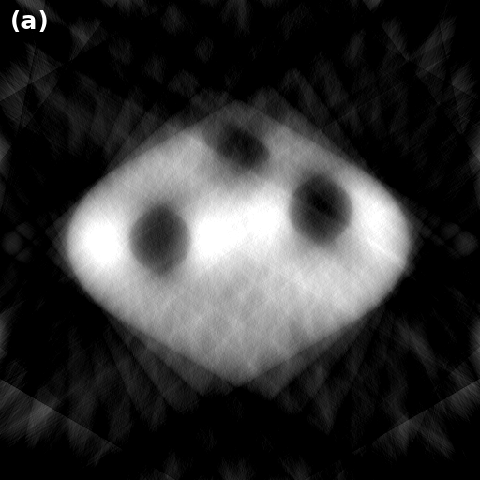}
				}
				\subcaptionbox{CS, 229s, RME: $0.0850$ \label{cs-simul-cs-comp-missing-wedge}}[.24\textwidth]{
					\includegraphics[scale=0.36]{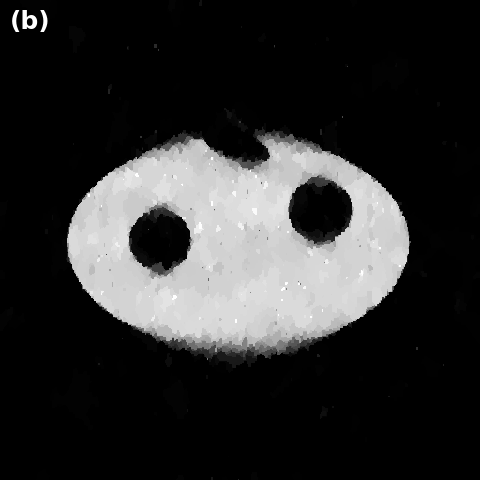}
				}
				\subcaptionbox{TVR-DART, 193s, RME: $0.2164$ \label{cs-simul-tvr-dart-comp-missing-wedge}}[.24\textwidth]{
					\includegraphics[scale=0.36]{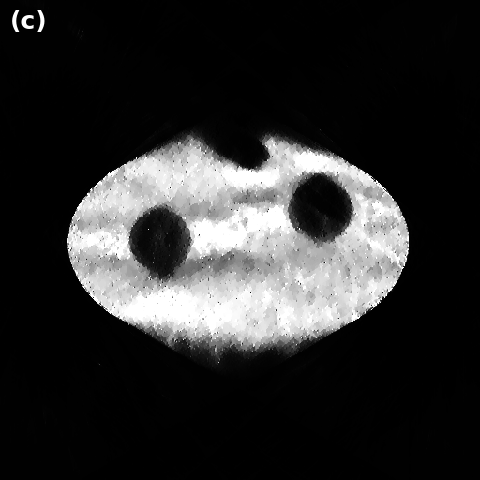}
				}	
				\subcaptionbox{CSHM, 69s, RME: $0.0495$ \label{cs-simul-cshm-comp-missing-wedge}}[.24\textwidth]{
					\includegraphics[scale=0.36]{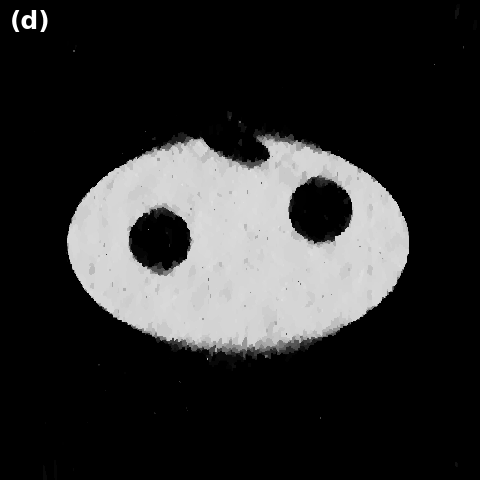}
				}
				\subcaptionbox{SIRT, 3s, RDC: $21.5546$ \label{cs-et1-sirt-comp-missing-wedge}}[.24\textwidth]{
					\includegraphics[scale=0.36]{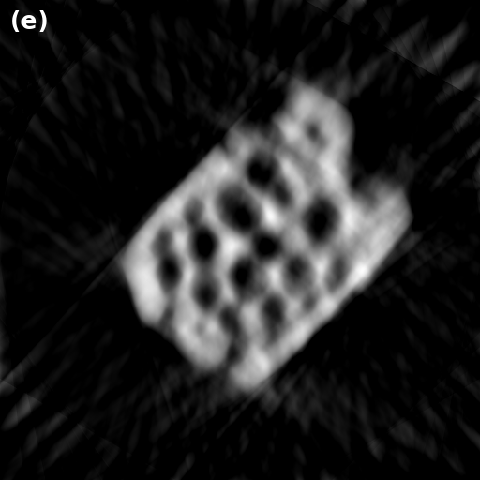}
				}
				\subcaptionbox{CS, 329s, RDC: $12.1703$ \label{cs-et1-cs-comp-missing-wedge}}[.24\textwidth]{
					\includegraphics[scale=0.36]{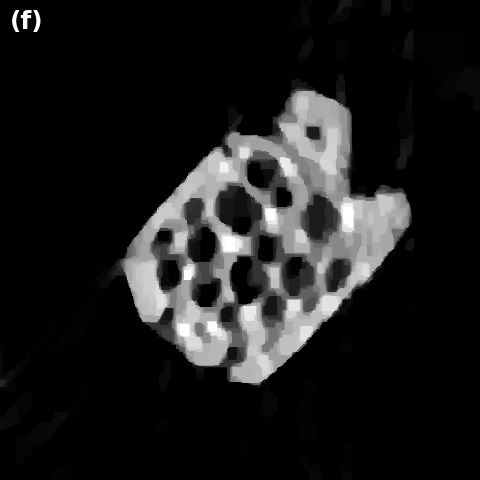}
				}
				\subcaptionbox{TVR-DART, 62s, RDC: $15.9070$ \label{cs-et1-tvr-dart-comp-missing-wedge}}[.24\textwidth]{
					\includegraphics[scale=0.36]{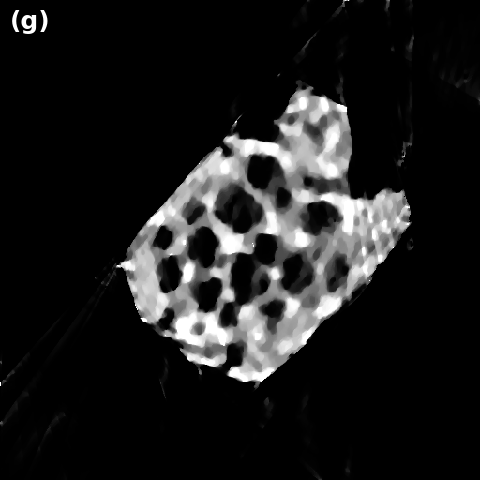}
				}
				\subcaptionbox{CSHM, 92s, RDC: $12.0015$ \label{cs-et1-cshm-comp-missing-wedge}}[.24\textwidth]{
					\includegraphics[scale=0.36]{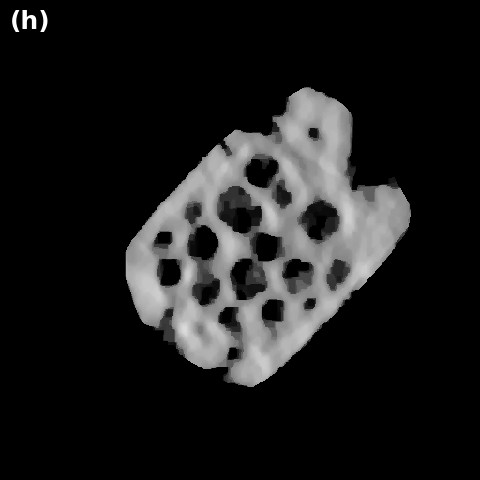}
				}
				\caption{Results for a missing wedge of $60\degree$ with $512^2$ pixels and $16$ projection angles for the simulated data set (a)-(d) and $21$ projection angles for the ET data set (e)-(h).}
				\label{fig:exp-comp-missing wedge}
			\end{figure*}

			In some nanotomography applications, the angular tilt range of the projections is limited due to, e.g., shadowing of the sample holder, leading to so-called missing-wedge reconstruction artifacts. \cite{leary2013compressed} In \figref{fig:exp-comp-missing wedge}, CSHM is compared to SIRT, CS, and TVR-DART on the simulated and the zeolite particle ET data sets with $512^2$ pixels and a missing wedge of $60\degree$. The respective projection angles are equidistantly spread over the remaining angular range of $120\degree$. Beneath the reconstructed slices, the corresponding RME/RDC values and running times are displayed.
			
			The magnitude of the running times of the algorithms are comparable to the cases without missing wedge. CSHM outperforms the other algorithms for the simulated data set (\figsref{cs-simul-sirt-comp-missing-wedge}-\ref{cs-simul-cshm-comp-missing-wedge}), both visually and regarding to the relative mean error. For the zeolite, CSHM also reaches the best RDC value, while being visually the best reconstruction. The typical elongation of reconstructions with missing wedge in vertical direction can be seen in both CSHM reconstructions of \figref{cs-simul-cshm-comp-missing-wedge} and \figref{cs-et1-cshm-comp-missing-wedge}, but the effect is relatively small, especially when compared to the SIRT reconstruction. In \figref{cs-et1-cshm-comp-missing-wedge}, the pore at the lower part of the zeolite is most likely a distorted and erroneously reconstructed indentation with merged borders, which can be deduced from a comparison with the $180$ angles SIRT reconstruction without missing wedge in \figref{ground-truth-image-comps}b. This merging of pore walls cannot be observed in the full tilt-angle range SIRT reconstruction, nor in the missing-wedge SIRT, CS, and TVR-DART reconstructions in \figsref{cs-et1-sirt-comp-missing-wedge}-\ref{cs-et1-tvr-dart-comp-missing-wedge}. However, in those reconstructions, the missing-wedge effect causes a much stronger erroneous merging of pores (instead of pore walls), which are separate in \figref{ground-truth-image-comps}b, compared to CSHM.


\section{Conclusion}
\label{sec:concandoutl}
        The previous section provided results obtained with the two newly developed constraints of this work. CS was combined with both hard and soft upper bounds, resulting in the here-called compressed sensing for homogeneous materials (CSHM) algorithm. CSHM was compared to SIRT, CS, and TVR-DART on simulated and experimental two-dimensional projection data from different nanotomography techniques, namely 360°-ET and nano-CT. Different problem settings were studied, including problems with a missing wedge.

		Regarding reconstruction quality, the results of CSHM are very convincing and of  high quality. In almost all cases for the simulated data set, CSHM achieved the lowest relative mean error and was only one time slightly outperformed by TVR-DART. CSHM also showed significantly decreased running times compared to CS, when solved with the same IPM. Both artifacts outside and inside of the reconstructed samples were reduced in comparison to CS, especially regarding the intensity of unexpected brighter spots inside and streaking artifacts outside of the samples. The artifacts outside of the samples were reduced by the hard upper bounds. Bounding variables to (near) $0$ lets the solver eliminate a lot of variables resulting in faster solving and less required memory than without the constraint. This effect relies on a preprocessing step of subtracting the mean background noise, to have a background intensity near $0$. The preprocessing step can be applied, if a mean background noise value can be estimated, e.g., if there are rays known that only pass through vacuum or air and do not hit the sample. The hard bounds are not only applicable to homogeneous samples, but to a much broader range of image reconstruction problems. They are especially efficient, when the sample is surrounded by vacuum or air. The artifacts inside the sample are reduced by the soft upper bounds, which are only applicable to homogeneous samples. They penalize larger deviations above the calculated homogeneous material density $\omega$, leading to the density of the reconstructed sample being closer to homogeneity. A slight drawback of CSHM was shown in form of a decrease in feature sizes, e.g.,\ pore sizes, by distributing exceeding material on the feature edges in the CSHM reconstructions. However, this effect is very small.
		
        An advantage of CSHM compared to TVR-DART is the consistency of good solutions. From a qualitative perspective, TVR-DART performs very well for some instances, possibly better than CSHM. However, for some other instances it returns bad image reconstructions, as can be observed in \figref{tvr-dart-comp-exp-1}. CSHM however, performs consistently well, even for very few projection angles.

		Missing-wedge problem reconstructions of CSHM show typical missing-wedge artifacts like elongation and the merging of features that are separate in reconstructions without missing wedge. Similar effects are observed for reconstructions of all the comparing algorithms. In comparison, CSHM exhibits the fewest merging of pores and an elongation similar to CS but slightly worse than TVR-DART.
		 
  Using more projection angles or a higher pixel count increases the running time for CSHM. Then, TVR-DART might become attractive with shorter running times, as the increment in quality of CSHM opposite TVR-DART might not be worth the large increment in running time anymore. However, as \tabref{result-table} indicates, the CSHM reconstructions do not improve by much anymore for an increasing number of projection angles, compared to a reconstruction with only a few angles, which is already very accurate. 
  This justifies reconstructing images with only a few projection angles, keeping the running time within an acceptable range. When reconstructing with only very little information (very few projection angles), CSHM outperformed TVR-DART in both running time and quality, as TVR-DART did not converge to a good reconstruction. Then, CSHM represents the most efficient choice for image reconstruction. To decrease high running times, existing work can be applied to the used IPM. Multiple approaches were proposed in the literature, e.g., using a matrix-free IPM to speed up the search step of the Newton method, developed by Fountoulakis et al.\ \cite{fountoulakis2014matrix}, or accelerating the IPM by performing parallel computations on the GPU, where promising results were obtained by Lee et al.\ \cite{lee2022gpu}.
		
		In conclusion, CSHM represents a competitive alternative to existing algorithms for smaller and medium-sized image reconstruction problems of homogeneous samples. It provides high quality results for highly undersampled data in nanotomography, including missing-wedge problems. Non-homogeneous sample reconstructions can still benefit from the hard upper bounds that reduce artifacts around the studied objects. Furthermore, an extension of the soft upper bounds to a version that can handle samples composed of few, but more than one, density values is thinkable. This is a question for future work.
\section*{Author Contributions}
Sebastian Kreuz: Methodology, Software, Investigation, Writing - Original Draft. Benjamin Apeleo Zubiri: Conceptualization, Methodology, Validation, Writing - Review \& Editing. Silvan Englisch: Validation, Data Curation. 
Sung-Gyu Kang: Writing - Review \& Editing. 
Rajaprakash Ramachandramoorthy: Writing - Review \& Editing.
Erdmann Spiecker: Writing - Review \& Editing.  Frauke Liers: Conceptualization, Methodology, Writing - Review \& Editing. and Jan Rolfes: Conceptualization, Methodology, Writing - Review \& Editing.

\section*{Data availability}
The tilt series raw data and the Python scripts of the implemented algorithms are publicly available on Zenodo \href{https://zenodo.org/doi/10.5281/zenodo.10283560}{DOI: 10.5281/zenodo.10283560}. 

\section*{Conflicts of interest}
 There are no conflicts to declare.

\section*{Acknowledgements}
We acknowledge funding by the Deutsche Forschungsgemeinschaft
(DFG, German Research Foundation) - Project-ID 416229255 - SFB 1411. We thank Alexander Götz for the help with the acquisition of the nano-CT data set.

\printbibliography

\end{document}